# Mesoscopic Origin of Ferroelectric-Ferroelectric Transition in BaTiO$_3$

Asaf Hershkovitz,[1,2] Florian Johann,[3] Maya Barzilay,[1,2] Alon Hendler Avidor[1,2] and Yachin Ivry[1,2,*]

[1]Department of Materials Science and Engineering, Technion – Israel Institute of Technology, Haifa 3200003, Israel
[2]Solid State Institute, Technion – Israel Institute of Technology, Haifa 3200003, Israel
[3]Asylum Research, an Oxford Instruments company, 6310 Hollister Ave, Goleta, CA 93117, U. S. A.
*ivry@technion.ac.il.

**Ferroelectric materials are the core of common technologies, such as medical ultrasound, mobile-phone antennae and low-power memory devices. The technological interest in ferroelectrics stems from the existence of switchable mesoscale polarization domains. Hence, understanding the origin of ferroelectric functionality requires realization of the domain dynamics during a ferroelectric transformation. However, domain dynamics characterization at the mesoscale is typically too slow with respect to the abrupt ferroic transition. Using scanning probe microscopy with 15-mK thermal-, and deep-submicron spatial- resolution, we realized the domain dynamics during an orthorhombic-to-tetragonal transition in the seminal ferroelectric BaTiO$_3$. We show that the transition comprises four distinguishable mechanisms. The dominant mechanism is a step-by-step progression of a tetragonal-domain wavefront into the orthorhombic phase. This progression is accompanied by ripple-like surface irregularities. Small island domains that remained orthorhombic diffuse then slowly after the wavefront progression. Finally, the resultant tetragonal domains equilibrate by coalescing in a constant-speed. These observations, which are accompanied by quantitative data, bridge between existing macroscopic and microscopic models regarding the nature of ferroelectric transitions, showing the mesoscale origin of ferroelectricity.**

After Helen Megaw observed experimentally[1] the symmetry change in $BaTiO_3$ already in 1945, Cochran[2] suggested that the transition is governed by lattice instability. The description of a spontaneous symmetry change as the origin of ferroelectricity has been adopted by Anderson[3] and Ginzburg.[4,5] In these models, higher-symmetry crystallographic phase transfers to the lower-symmetry phase (or vice versa) in an organized manner, as in a card-house collapse. That is, the transition is displacive and is governed by collective non-local interactions.[6] Displacive transitions are diffusionless,[7] which are not very common and exist, e.g. in some ferromagnets and superconductors, mainly those that exhibit ferroelasticity.[8] The displacive mechanism in ferroelectrics was then questioned when x-ray scattering results implied[9] on short-range interactions, which give rise to a diffusive transition at random points, as in the case of bubble formation in boiling water. This diffusive mechanism is different from an organized displasive transition. Later first-principle[10] and computer-based[11] models suggested that the onset of ferroelectricity is a displacive-diffusive mixture. These predictions were then supported by macroscopic x-ray[12] and magnetic-resonance[13] characterizations. Nevertheless, the origin of the transition at the domain scale has remained elusive.

Submicron-resolution studies of domain statics are commonly available, e.g. with piezoresponse force microscopy (PFM).[14] Some works involve also static domain imaging of several different crystallographic phases.[15–17] Dynamic characterizations are more challenging due to the typical short time scale of the transition with respect to the measurement time.[18,19] Only handful successful mesoscale domain-dynamic observations have been reported,[20–22] mainly during electric-field switching. However, a combination of high-resolution and rapid imaging of mesoscale domain dynamics at variable temperatures is crucial for understanding the origin of the ferroic transition, while such measurements are unavailable.

Ferroelectric-ferroelectric transitions are of a special interest for realizing the domain organization not only during, but also before and after the phase transformation. Observing domain dynamics around such a transition allows an excellent opportunity for determining the kinetics of ferroic domains.[23–29] Here, we revealed the mesoscale domain dynamics at the orthorhombic-to-tetragonal transformation in a single crystal $BaTiO_3$ by means of variable-temperature PFM mapping with high temperature stability. Simultaneous topography mapping allowed an important complementary characterization. The temperature stability allowed us to rump up the temperature reliably during the phase transition by less than 1º C over the course of ~48 minutes, slowing down the transition to the time scale relevant for PFM imaging. We demonstrated that the transition takes place in four distinguishable steps: (i) pre-transition material softening; (ii) displacive tetragonal wavefront penetration towards the orthorhombic phase with local order-disorder variations; (iii) diffusive pinning-depinning of domain islands; and (iv) post-transition zipping-like domain-coalescence relaxation of the resultant phase. We demonstrate the dynamics of each mechanism by means of direct observations of the domain evolution, available as both videos and figures. Moreover, we quantified the typical lengthscale and time scale of each mechanism. We show that each of the four steps complies with a different theoretical prediction, so that competing models describe the constituent steps of the phase transition. Finally, the coexistence of tetragonal-orthorhombic domains supplies us with a direct observation of the transition's first-order nature.

The primary part of the transition occurred when a zigzag wavefront of striped tetragonal domains penetrated the orthorhombic domain mesh with a nearly constant speed of 32 nm s$^{-1}$. Video SI1 and Figure1 show the wavefront progression in a $86 \times 86$ μm$^2$ during 2884 s, when the $BaTiO_3$ was heated gradually by 0.97⁰ C.[30] The angle between two zigs in the zigzag wavefront is roughly 11º, while the angle between the orthorhombic and tetragonal domains is 45º. Following

Marton et al.,[31] the observed 45⁰ orientation of orthorhombic domain walls with respect to the pseudo-cubic structure as well as the zigzag shape of the wavefront (11⁰ opening) indicate that domains slide along the (0.7,0.13,-0.7) plane (see Extended Data Figures 1 and 2 for a more detailed crystallographic analysis). The continuous wavefront progression of the sliding orthorhombic-tetragonal domain walls along a well-defined crystallographic plane indicates that the dominant part of the transition is of a displacive. Moreover, the coexistence of tetragonal and orthorhombic domains during the ferroelectric transformation clearly indicate on a first order transition.

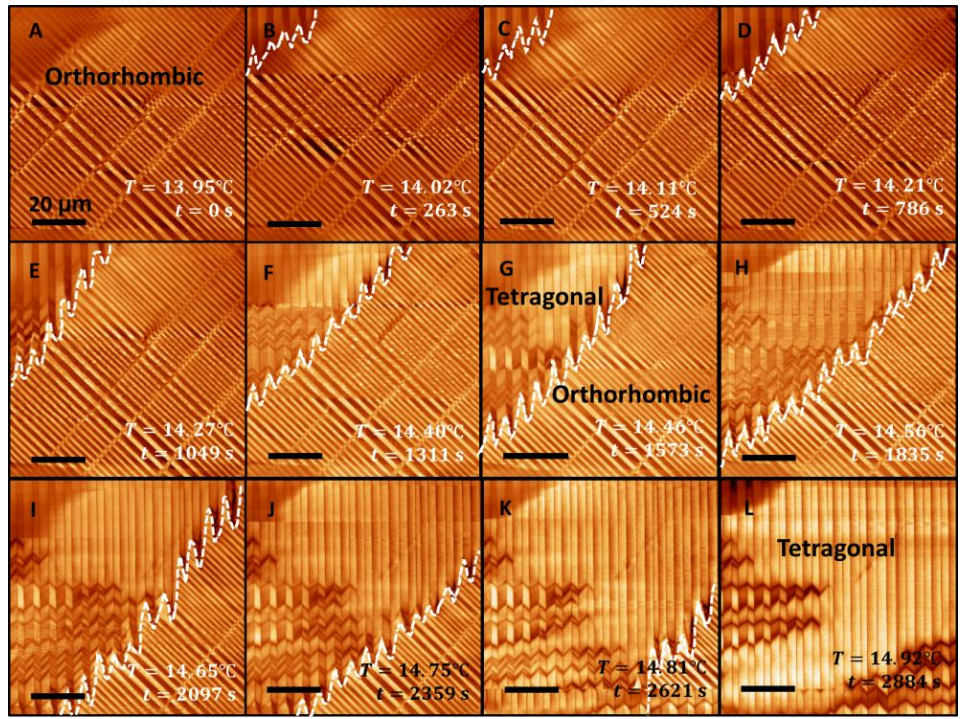

**Figure 1 | Ferroic domain dynamics during orthorhombic-to-tetragonal transition: wavefront progression**. Upon heating a BaTiO$_3$ single crystal, orthorhombic elastic domains (**A**) were gradually being pushed away by a progressing tetragonal wavefront (highlighted with a white curve). (**B**-**K**) This transition occurred by a sliding tetragonal-orthorhombic domain wall along the $(0.7, 0.13, \overline{0.7})$ pseudo-cubic plane (see Figure SI1) until a uniform tetragonal striped structure

finally dominated (**L**). The tetragonal cusp of the wavefront penetrated further the tetragonal *c* domains (darker stripes), while the *a*-domain (brighter stripes) cusp was predominantly orthorhombic (see Figure SI1-2). Local zigzag shape variations within the larger scale wavefront indicate on the mixed short- and long- range dynamics at the transition. Temperature (*T*) and time (*t*) of the lateral PFM amplitude scans are overlaid.

The wavefront progression governs the transition from orthorhombic to tetragonal in most of the material. However, in some areas, the domains undergo the transformation with a competing mechanism that takes place in parallel to the wavefront progression. Two effects can be attributed to this competing switching mechanism, which requires a more careful treatment. First, there are some local variations in the ~11° zigzag opening angle (Figure 1). Secondly, when the variations in the zigzag shape are large, the striped tetragonal domains not continuous after the wavefront had passed. Rather, small island-like bundle domain[32] structures seems to nucleate at the points where the wavefront was distorted (See Figure 2 as well as in [Video SI2](Video SI2)). Following previous analyses,[31] these islands are most likely not tetragonal, but are remnants of orthorhombic domains or an intermediate monoclinic structure.[16,33,34]

The correlation between the formation of island domains and strong variation in the orthorhombic-tetragonal common domain wall (i.e. distortion of the wavefront) suggests that the creation of these domains is associated with high local strain concentrations. Moreover, as opposed to the rather homogeneous nature of the wavefront-progression transition, the island-domain distribution is inhomogeneous. Such inhomogeneous local strain variations imply that island domains are formed due to pinning sites, which are induced either by local crystallographic defects in the native crystal or due to local strain concentration that evolves during the transition. A possible explanation to the origin of the latter is given in Figure SI2.

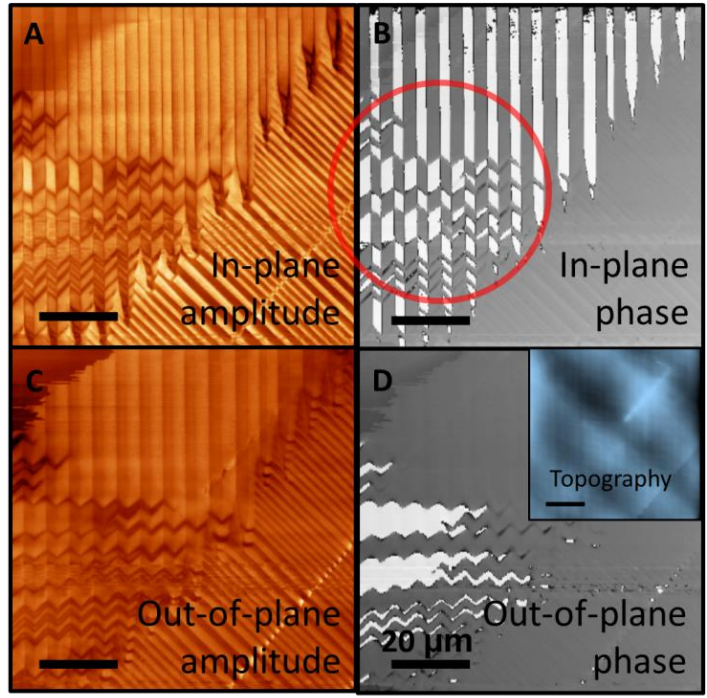

**Figure 2 | Island domain formation during a ferroic transition.** During the wavefront progression (Figure 1), erratic domain islands were formed (highlighted). Simultaneously measured signals of PFM in-plane (**A**) amplitude and (**B**) phase, PFM out-of-plane (**C**) amplitude and (**D**) phase (insert: topography) at $T = 14.65^0$ C and $t = 2097$ s (corresponding to Figure 1i) allowed us to analyze[31] that these domains are not tetragonal (rather, they are most likely either remnants of orthorhombic domains or an intermediate monoclinic structure[16,33,34]).

The island domains complete the transition to a continuous tetragonal striped domain structure in a much slower process in comparison to the displacive transition of the wavefront progression. Video SI2 and Figure 3 shows that this depinning relaxation process took place even when the temperature was held constant already above the transition ($T = 20 \pm 0.015^0$ C). We found that the change in area size was rather linear with time. Although each island domain had a different relaxation rate, we calculated the fastest relaxation speed to be $< 4000 \pm 300$ nm$^2$ s$^{-1}$ (see Figure 3F for details). The observed slow relaxation and arbitrary location of the islands indicates that this transformation mechanism is not displacive. Rather, the observed pinning-depinning mechanism is of an order-disorder diffusive nature. This mechanism that has been explained theoretically[35] and was confirmed with previous x-ray diffraction experiments[33] is supposedly competing with the displacive transition. However, the observation of both a displacive (wavefront progression) and diffusive (islands) mechanisms that take place at the same substance, supplies a direct evidence that these two competing mechanisms can coexist, albeit at different length and time scales.

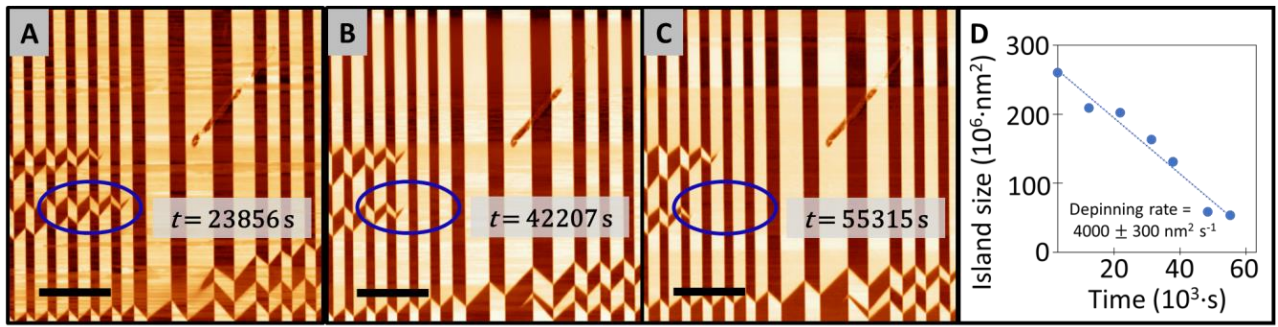

**Figure 3 | Island domain depinning relaxation.** (A-C) Lateral PFM amplitude images of the island domain depinning at $T = 20 \pm 0.015^0$ C (blue oval helps guide the eye of the fastest process). All scale bars are 20 μm. The detailed depinning relaxation is given in Video SI2. (**D**) Representative data of the domain island depinning dynamics which is given by the dependence of the island area on time, showing $< 4000 \pm 300$ nm$^2$ s$^{-1}$ relaxation rate, which is much slower than the displacive wavefront progression, indicating on the diffusive nature of this process.

The wavefront progression and island depinning are the only two mechanisms we identified as in charge of the orthorhombic-to-tetragonal domain transformation. Yet, the transition is accompanied by clear pre-, and post- transition effects that must also be taken into account when discussing the ferroic phase transformation. These effects seem to be essential for the formation and completion of the transition and can be attributed to releasing effectively excess strain in the crystal. Thus, although each effect involves merely one type of domains—orthorhombic or tetragonal, the pre- and post- transition mechanisms have to be discussed within the framework of the ferroic phase transformation.

Figure 4 illustrates the surface topography of the $BaTiO_3$ throughout the wavefront progression part of the transition (Figure 1). The data (Figure 4a) show clearly that prior to the transition, the topography included quasi-periodic hills. These wrinkles that are designated in Roman numerals are 2.5-12-nm high and are of a double periodic structure of a short (I-III and IV-V, ~15 μm) and long (I-II and III-IV, 20-25 Roman μm) spacing. Figure 4 and [Video SI1](#) demonstrate that these quasi-periodic wrinkles[26] exist in the orthorhombic phase, just before the wavefront arrives. As the wavefront progresses, wrinkles next to the wavefront (but yet at the native orthorhombic phase) dissipated, while new wrinkles appear further away. This disappearance and appearance of the quasi-periodic wrinkles looks similar to a lateral sliding of a line fold in a paper sheet, only here the wrinkle motion is discrete. The wrinkles (topographic corrugation) and wavefront location (adopted from the PFM data of Figure 1) are both highlighted in Figure 4, where the disappearing and appearing wrinkles are also illustrated (dashed and continuous lines, respectively, in the schematic inserts). The quasi-periodic organization of the wrinkles and the wrinkle dynamics imply on a ripple-like behavior that is common for granular or viscous media (e.g. bedforms[36]).

That is, at the pre-transition state, the material softens. Such softening can be associated with a relaxor-like behavior, which is common in ferroelectrics that their transition spans a temperatures range that is much broader than the transition reported here.[46] This softening is also reminiscence of the concept of Cochran's early model.[6]

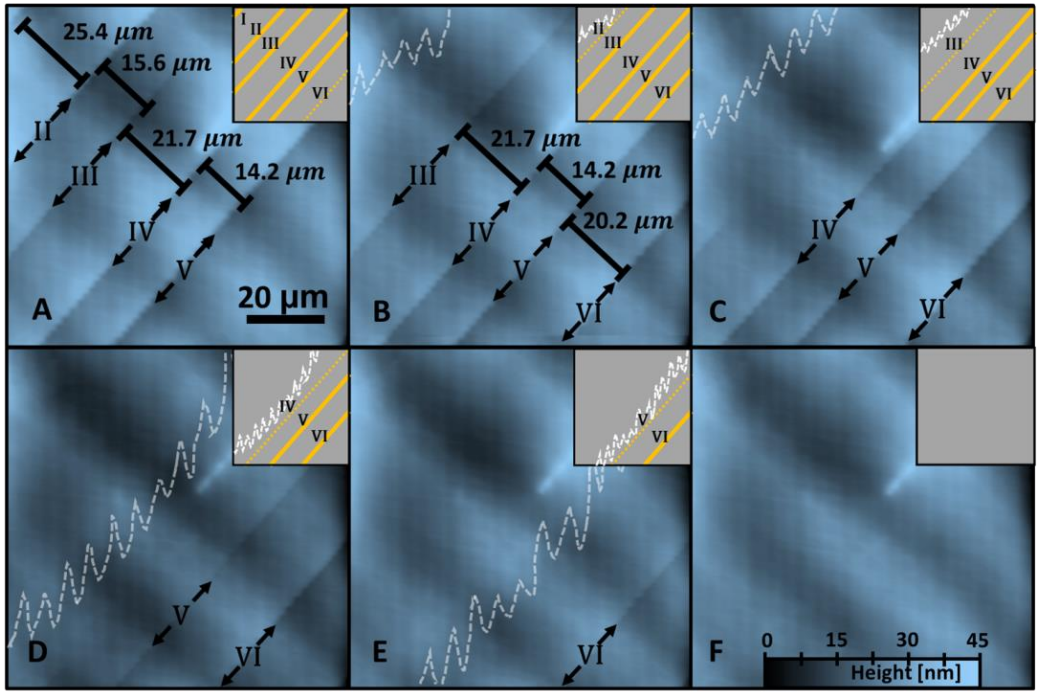

**Figure 4 | Surface irregularity dynamics at the pre-transition state.** Topography imaging of the BaTiO$_3$ reveals (**A**) six initial wrinkles at the orthorhombic phase (I-VI). These surface irregularities were organized in alternating short (I-III and IV-V) and long (I-II and III-IV) quasi-periodic spacing and are ca. 10-nm high. Upon heating, the striped tetragonal domains gradually grew from top to bottom (see Figure 1), displacing the wrinkles (zigzag-shaped wavefront that is adopted from the simultaneously taken PFM images in Figure 1 is highlighted). First, (**B**) the wrinkle 'I' disappeared and a new wrinkle, 'VI,' appeared (~20.2 μm away from 'V'). Next, (**C-E**) the wavefront progressed and the wrinkles gradually dissipated, serving as a precursor of the wavefront position. Finally (**F**) the entire area became striped and no wrinkles were seen. Here,

temperature and time (following Figure 1) are $T = 13.95, 14.02, 14.21, 14.46, 14.75, 14.92^0$ C and $t = 0, 263, 786, 1573, 2359, 2884$ s for the corresponding (A-F) images of the same area. The complete wrinkling dynamics is given alongside the PFM images that show contain wavefront progression in Video SI1.

In addition to the pre-transition softening effect, our observations clearly indicate on a post-transition strain-releasing mechanism of the newly formed tetragonal domains. Towards the end of the transition, after the wavefront had passed, another relaxation took place in the striped tetragonal domains. Here, two neighboring *a* stripes coalesced and wiped out the *c* domain that was sandwiched between them, so that a broad single *a* domain was formed. The domain coalescence took place in a zipping-like process. By tracing the progression of the zipping apex, we found that this process is linear in time (i.e. constant speed). We found that there are usually several sequential zipping events (see Video SI3). Despite the linearity in time of each zipping event, the speed decreased gradually between the sequential events from about 20 nm s$^{-1}$ to 2 nm s$^{-1}$. The gradual decrease in zipping speed indicates on a gradual release of remnant stress. Figure 5 shows representative data from the slowest zipping events. The complete data from the zipping events as well as quantitative analyses of their linear rates are given in Video SI3.

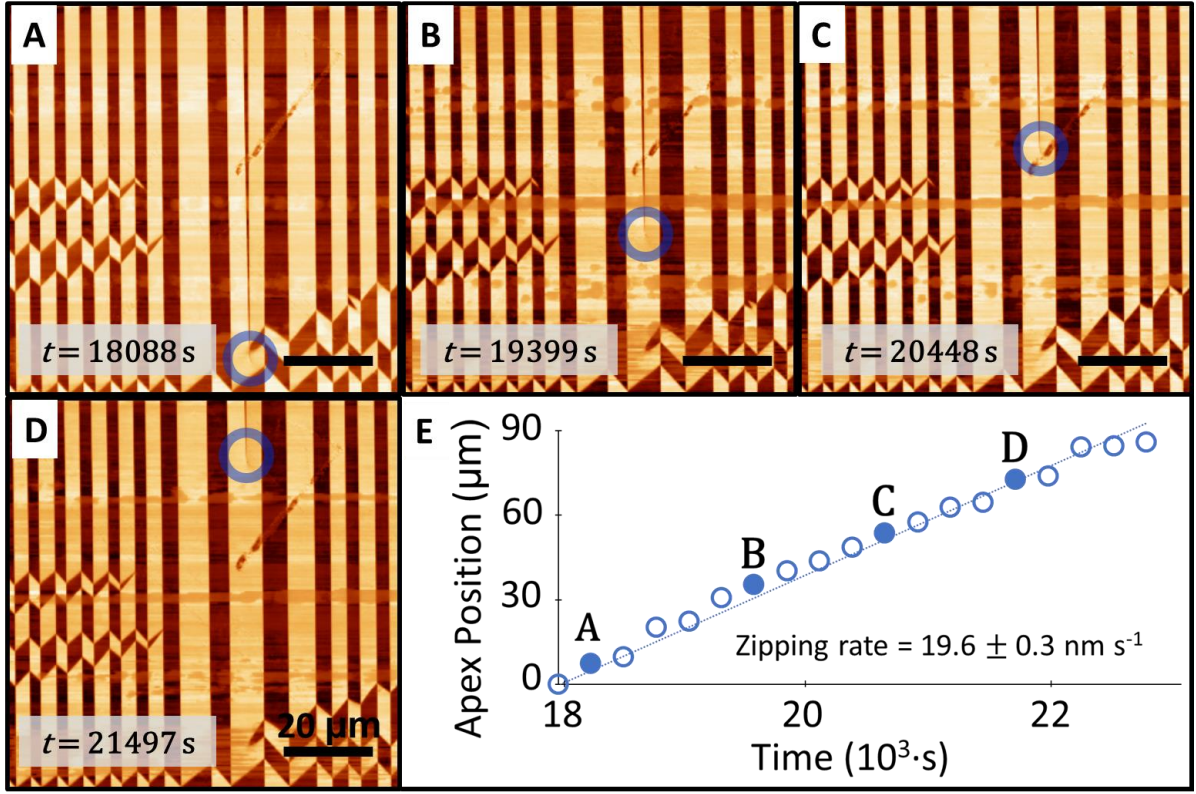

**Figure 5 | Constant-speed zipping-like coalescence of tetragonalstriped domains.** (**A-D**) Striped elastic tetragonal domains merged into broader stripes in a zipping like process (arrows help guide the eye). (**E**) Zipping progression as a function of time demonstrates a constant speed: $19.6 \pm 0.3$ nm s$^{-1}$. Data points obtained from the zipping locations in (A-D) are labeled. The complete data and additional zipping events are given in Video SI3.

The zipping process exposes the nature of domain dynamics at the nearly homogeneous post-transition tetragonal structure. Domain dynamics of tetragonal stripes play a significant role in modern ferroelectric-based technologies.[20] Striped-domain coalescence in the form of domain doubling has garnered attention recently,[26,32] but real-time dynamics realization is still absent. Previous studies suggest that domain mobility is dominated by creep,[25] i.e. local crystallographic defects[38] that diffuse with a power-law rate. Although our observations of a constant-speed zipping do not support such a description, typically, the power-law exponent is of the order of unity,[25,39] namely, a near-constant speed. Yet, the constant speed mobility is rather unexpected and does not comply with the $-\frac{1}{3}$ time exponent that the Lifshitz-Slyozov model[40] predicts, when accommodating the process of Ostwald ripening to supersaturated solid solutions. A possible explanation is that while each of the zipping events is of a constant speed, the speed varies between sequential zipping events. Thus, at the mesoscopic scale, the inhomogeneous domain behavior that is often addressed in other studies, averages and does not show a constant speed at the macroscopic domain mobility.

We should note that the zipping relaxation may arise to release strain in the post-transition phase, e.g. for complying with Roytburd's scaling law.[41] Strictly speaking, the effect is not completely post transition because some non-tetragonal islands still exist elsewhere. Likewise, the topography ripples in pre-transition softening state are observed in the region that contains only orthorhombic domains, i.e. they are taking place before the transition and can be referred to as the precursor or the miners' canary of the transition. However, the wrinkles occur when the transition already began in a neighboring area and some parts of the crystal was already transferred to tetragonal. Thus, both the pre-transition softening and the post-transition zipping may be considered as parts of the transition, allowing us to distinguish between the four different mechanisms spatially. However, when characterizing the transition macroscopically, one cannot distinguish between processes that take place in the tetragonal domains (zipping), processes that take place in the orthorhombic domains (wrinkles) and processes that involve both phases (wavefront progression and island domain depinning). Therefore, the direct observation of the domain dynamics during (as well as before and after) the transition supplied here may help explain some of the contradictions in the literature regarding the dispute over the displacive versus diffusive nature of the transition.

The four-step transition reported here indicates on the mesoscale domain origin of a displacive-diffusive nature of the ferroic transition as well as of the long-range (wavefront) and short-range (i.e., zigzag deformations and depinning of the island domains as well as the post-transition zipping like domain coalescence, respectively) competition it encompasses. Our observations thus complement for the commonly agreed three–step process domain dynamics: nucleation, linear motion of needle-like domains and sideways growth of domains[23,25,42] that presumably dominate domain switching as well as the paraelectric-to-ferroelectric transformation. We should note though that with respect to domain nucleation, it is plausible that the wavefront reported here originated from a nucleation point somewhere else in the crystal. Future studies that aim at tuning the interplay between the entire or part of the four transition steps seen here, e.g. by introducing electric field, stress or controlled environment may help prioritize one of the four mechanisms over the others, perhaps leading to improvement in the functionality of ferroelectrics or other ferroelastic functional materials.[43]

**Methods**

We used single crystals BaTiO3 of 5×2.5×1 mm in size. Samples (three) were obtained from both MaTecK and SurfaceNet, while the samples were acetone and ethanol cleaned. X-ray diffraction was done using Rigaku SmartLab diffractometer with Cu electrode ($\lambda = 0.15406\ nm$) and $\theta - 2\theta$ measurement (Figure SI4). The samples were scanned both in a fast broad-spectrum scan (5°-90°, 7° min$^{-1}$ and 0.01° jumps) as well as in a slower closer-look scan (44°-46.5°, 1° min$^{-1}$ and 0.01° jumps).

X-Ray photoelectron spectroscopy (XPS) measurements were performed in a PHI – Versaprobe III microprobe system (PHI, USA). XPS surface analyses were done with a monochromatic Al (Kα) radiation of 1486.6 eV energy and emitted photoelectrons were directed to a spherical capacitor analyser (SCA). The x-ray beam created at the surface of the sample localized positive charges. In order to palliate to this localized surface charge, we used the PHI's dual-beam neutralization technique. Photoelectron take of angle (TOA) was 45° and the photoelectron collection angle was ±20°. The analysis area was 200 μm in diameter. Survey spectrum was recorded before and after sputtering the BaTiO3 single crystal with a 3-kV Ar$^+$ ion gun (data shown here in Figure SI4 were collected prior to the Ar$^+$ sputtering). The survey spectra were collected at the middle of the sample with a pass energy of 140 eV and a step size of 0.025 eV, from which the surface chemical composition was determined (Figure SI4a). The core level binding energies of the different peaks were normalized by calibrating the binding energy for C1s at 284.5 eV. The atomic concentrations were calculated using elemental relative sensitivity factors from Multipak Software library. High energy-resolution measurements were performed with a pass energy of 55 eV (Figure SI4b-d) and a step size of 0.05 eV.

Both x-ray diffraction and x-ray photon spectroscopy confirmed the high quality of the BaTiO$_3$ single crystals, while the native-domain PFM imaging (e.g. Figure 1a) revealed long-range ordered coherent ferroelastic domain organization, which implies on the high quality structure. PFM measurements were carried out with Asylum MFP-3D AFM and DCP01 Probe from TipsNano at both ambient and environmental control (with a constant flow of dry air in a sealed cell), deflection setpoint: ca. 100 nN. The images were taken in vector mode at vertical frequency of 860 kHz and lateral frequency of 1.38 MHz. Drive amplitude for both vertical and lateral excitations was 10 V. Fifteen orthorhombic-to-tetragonal transitions were imaged. All mechanisms have been observed at least twice, while some of the observed transitions took place too fast to allow inter-transition imaging (depending also on the heating rate that was changed from one experiment to another). Successful measurements were carried out also with a diamond tip (Adama, 1.56 N/m) while driving 15 V ac voltage at 197 kHz for the vertical signal and 6 V ac voltage at 603 kHz for the lateral signal. All PFM and AFM images were taken at 256×512 pixel-resolution scans, using the Asylum Research's software. For all images, both the backward and forward scans were recorded, while here, only the forward-direction images are presented for the sake of consistency. No manipulation has been conducted to the AFM and PFM images other than the usual AFM flattening that is compensating for the tip scan. Supporting Videos 1-3 include negligible contrast adjustments that compensate for the random contrast offset that the microscope produces naturally between images.

**Acknowledgments:** We would like to thank Prof. James F. Scott as well as Prof. Stanislav Kamba for helpful discussions. We would like to acknowledge Dr. Maria Koifman and Dr. Kamira Weinferd-Cohen for their support with the XRD and XPS experiments, respectively. We acknowledges financial support from the Zuckerman STEM Leadership Program, the Russel Barry Nanoscience Institute, Eliyahu Pen Research Fund as well as from the Israel Science Foundation (ISF) grant # 1602/17.

**Supporting Information:**

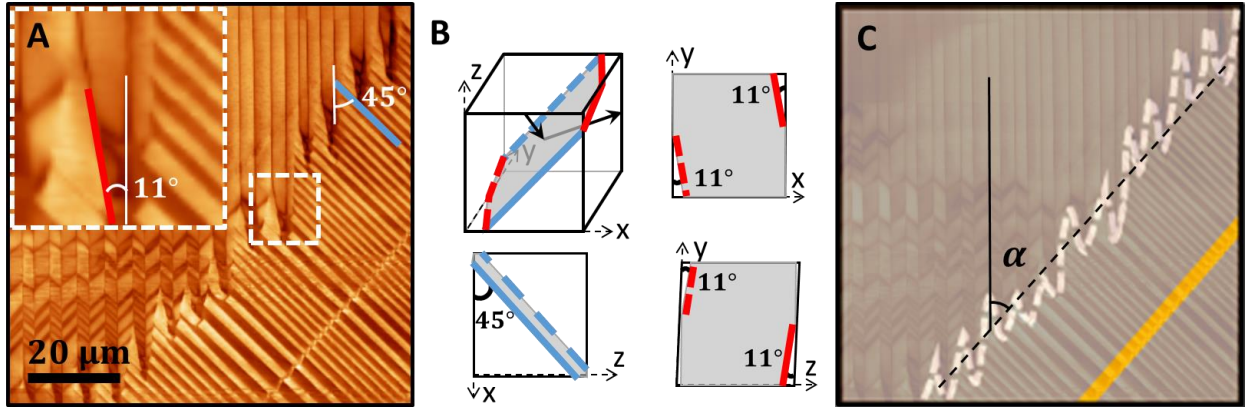

**Figure SI1 | Crystallographic analysis of the displacive-diffusive transition.** (**A**) A closer look at the tetragonal-orthorhombic border (taken from Figure 1I) shows 45⁰ orientation of the orthorhombic domain walls with respect to the pseudo-cubic structure. The cusp of the tetragonal *c* domain within the zigzag wavefront is ~11⁰. (**B**) Such a configuration corresponds to the energetically preferred O60 domains calculated by Marton et al.,[44] giving rise to 60⁰ polarization orientation (highlighted with arrows). Here, the domains slides along the twin plane of $(0.7, 0.13, \overline{0.7})$. Domain gliding along this plane is projected to a top-view angle of ~11⁰ wall, in agreement to the insert in (A). (**C**) The local crystallographic structure complies with the larger scale phase boundary (designated by $\alpha$), which is also ~45⁰. The closer look at (A) illustrates also that although the *c*-domain cusp is tetragonal, the *a*-domain cusp is still orthorhombic, indicating on a variation in transition rates for the different crystallographic orientations.

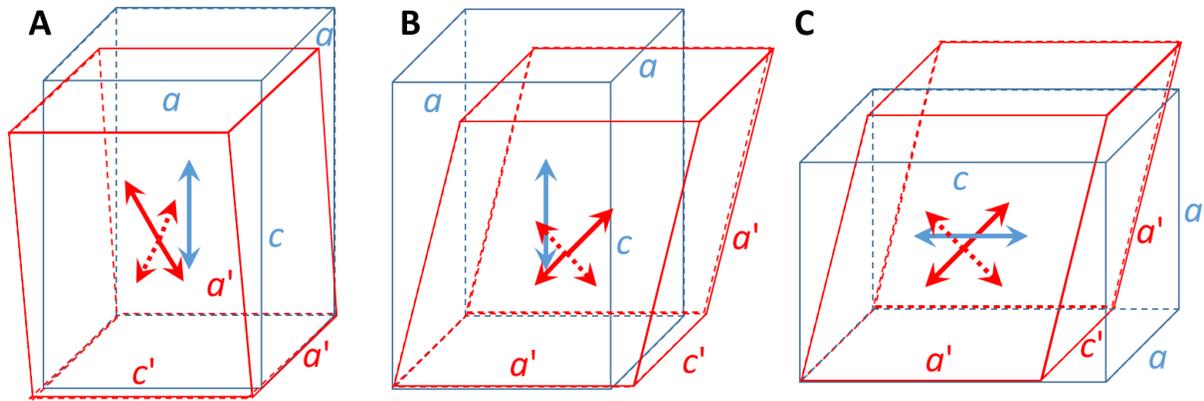

**Figure SI2 | Orthorhombic-to-tetragonal domain transformation.** Given the crystallographic configuration in Figure SI1, there are two possibilities (both have a two-fold elastic symmetry and four-fold electric symmetry) for the orthorhombic crystal to transfer to a tetragonal *c* domain (**A-B**), and only one possibility (also with a two-fold elastic symmetry) to transfer to a tetragonal *a* domain (**C**).

We propose that the higher probability to transfer to a tetragonal *c* domain is translated here to a faster progression of these domains highlighted in Figure SI1. Bearing in mind the orthorhombic-tetragonal configuration of the framework of Marton et al.,[31] there are eight options for the orthorhombic phase to transfer to a tetragonal *c* domains and only four to transfer to an *a* domain. We propose that this difference in probability is translated into a variation in transformation speed. This proposal is supported by the observation of a tetragonal cusp in the tetragonal *c*-domain region of the wavefront and an orthorhombic cusp at the *a* domains (Figure 1 and Figure SI1). This speed difference may add to the strain imposed by native defects, contributing to the observed local and temporary variations in the opening angle around the 11º. We suggest that this variation in speed which is either due to the native defects or due to the wavefront dynamics gives rise to a shear strain, which in turn produces pinning sites or a nucleation platform for the non-tetragonal island domains that are discussed in Figure 3.[35,45] Crystallographic orientations of the orthorhombic structure are highlighted with a' and b' with respect to the quasi-cubic structure, while the tetragonal orientations are designated with a and c. Therefore, although the wavefront progression seems to be the dominant mechanism of the unperturbed crystal, the island domains are of a diffusive nature that stems from local defects rather than due to a displacive transition along a permitted crystallographic gliding plane.

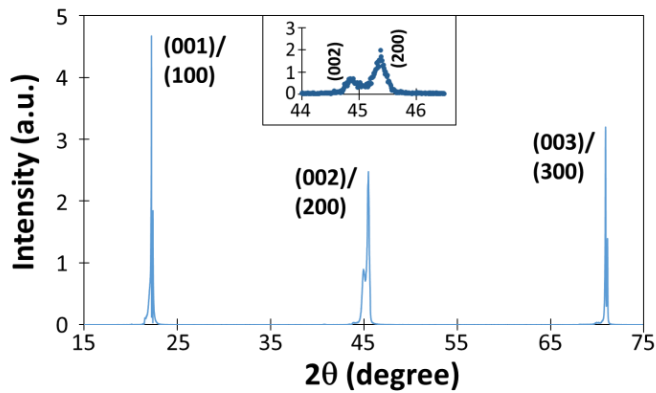

**Figure SI3 | X-ray diffraction of the BaTiO₃ at room temperature.** Broad-spectrum measurement reveals the existence of a BaTiO₃ single crystal only with peaks of the three first orders of the Bragg diffraction. Inset: high-resolution measurement around the second-order diffraction indicating on a clear separation between the out-of-plane and in-plane ferroelastic domain signals, corresponding to 0.398 nm and 0.402 nm of the *a* and *c* lattice parameters, and in agreement with the literature, including the seminal measurements of Megaw.[1,46]

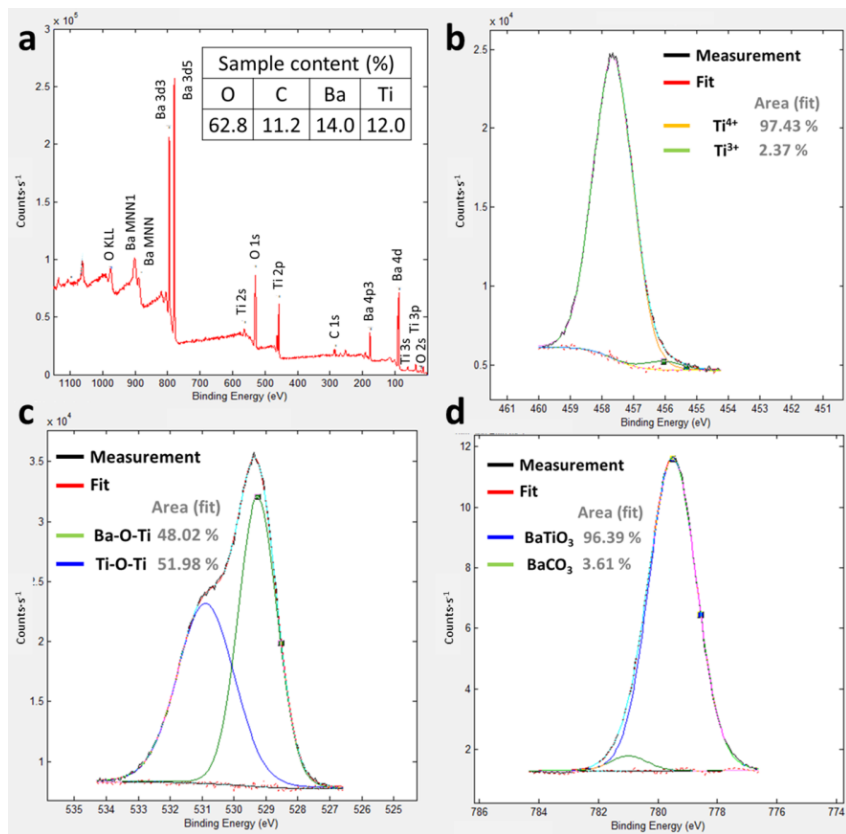

**Figure SI4 | X-ray photoelectron spectroscopy of the BaTiO$_3$ single crystal.** (**a**) Broad spectral survey. A closer look at the (**b**) oxygen, (**c**) BaTiO$_3$ and (**d**) titanium peaks reveals the purity of the single crystal. Small percentage of carbon surfactants exist on the native sample.

Videos SI1-3

[Video SI1](LINK) | **Wavefront progression and rippling at the phase transition ([LINK](LINK)).** Time evolution of the tetragonal wavefront that penetrates the orthorhombic phase (Left) observed with amplitude lateral PFM alongside topography wrinkling at the adjacent orthorhombic phase (Right). Time follows Figure 1.

[Video SI2](LINK) | **Depinning relaxation of pinned island domains ([LINK](LINK)).** Time evolution of depinning relaxation of island domains that were pinned after the wavefront had passed, showing a much slower process than the wavefront progression ([Video SI1](LINK)). Time follows Figure 1.

[Video SI3](LINK) | **Zipping of striped tetragonal domains ([LINK](LINK)).** Time evolution of three sequential zipping relaxation events in striped tetragonal domains (Left- amplitude lateral PFM) as well as of the measured location of the zipping apex after the wavefront completed progressing in the area. The characterized kinetics (Right) show that each zipping event is of a constant speed, while the speed decreases between the sequential domain coalescence events. Representative data from the first zipping event are presented in Figure 5. Time follows Figure 1.